\documentclass[a4paper,11pt]{article}
\pdfoutput=1 

\usepackage{jheppub} 

\usepackage[T1]{fontenc} 

\usepackage{float}
\usepackage[caption = false]{subfig}
\usepackage{ulem}
\usepackage{xcolor}
\newcommand{\LL}{\mathcal{L}}       

\newcommand{\OO}{\mathcal{O}}       

\newcommand{\pz}{\ensuremath{\sigma^z}}

\newcommand{\cdj}{c^\dagger_j}

\title{\boldmath On Complexity and Duality}

\author[a,b]{Jeff Murugan,}
\author[a]{Zayd Pandit}
\author[a,b]{and Hendrik J R Van Zyl}

\affiliation[a]{The Laboratory for Quantum Gravity \& Strings, Department of Mathematics and Applied Mathematics, University of Cape Town, Cape Town, South Africa}
\affiliation[b]{The National Institute for Theoretical and Computational Sciences, Private Bag X1, Matieland, South Africa}

\emailAdd{jeff.murugan@uct.ac.za}
\emailAdd{PNDZAY001@myuct.ac.za}
\emailAdd{jaco.vanzyl@uct.ac.za}

\abstract{We explore the relationship between complexity and duality in quantum systems, focusing on how local and non-local operators evolve under time evolution. We find that non-local operators, which are dual to local operators under specific mappings, exhibit behavior that mimics the growth of their local counterparts, particularly when considering state complexity. For the open transverse Ising model this leads to a neat organisation of the operator dynamics on either side of the duality, both consistent with growth expected in a quadratic fermion model like the Kitaev chain.   When examing periodic chains, however, the mapping of boundary terms provides access to multiple branches of highly complex operators. These give rise to much larger saturation values of complexity for parity-mixing operators and are in contrast to what one would expect for a quadratic Hamiltonian.  Our results shed light on the intricate relationship between non-locality, complexity growth, and duality in quantum systems.}

\begin{document} 
\maketitle
\flushbottom

\section{Introduction}
\label{sec:intro}
Understanding how operators evolve and grow in quantum many-body systems has become central to the study of quantum complexity and chaos \cite{Maldacena:2015waa, Hashimoto:2017oit, Roberts:2018mnp,Parker:2018yvk}. One promising approach to quantify this growth is through Krylov complexity \cite{Parker:2018yvk,Barbon:2019wsy,Dymarsky:2019elm,PhysRevE.104.034112,Rabinovici:2020ryf, Caputa:2021sib,Balasubramanian:2022tpr,Nandy:2024htc}, which measures how an operator spreads within its Hilbert space under time evolution. Krylov complexity is determined by the Lanczos coefficients, which describe the recursive application of the Hamiltonian on a chosen initial operator \cite{Parker:2018yvk, viswanath2008recursion}. This framework provides a powerful lens for investigating dynamical properties such as operator growth, scrambling, and quantum chaos \cite{Ali:2019zcj,Bhattacharyya:2020kgu,Bhattacharyya:2020rpy,Bhattacharyya:2020art,Bhattacharyya:2020qtd,Rabinovici:2020ryf, Bhattacharjee:2022vlt,Nandy:2024htc}. The systematic growth of Krylov complexity reveals distinct signatures in integrable and chaotic systems, with integrable models typically exhibiting polynomial growth, while chaotic systems show faster, often exponential, operator spreading \cite{Rabinovici:2020ryf,Bhattacharya:2023zqt, Sanchez-Garrido:2024pcy, Avdoshkin:2022xuw}.\\

\noindent
In this work, we explore Krylov complexity in the context of theories related under a duality transformation, testing the hypothesis that dual operators exhibit similar growth in Hilbert space, even when one operator is local and the other is non-local. The idea that non-local operators (those dual to local operators through bosonization or fermionization) could exhibit growth patterns akin to local operators is intriguing, particularly in the context of state complexity, which measures the difficulty of preparing quantum states from a reference state \cite{Balasubramanian:2022tpr}. Our primary goal is to understand how duality affects the operator growth and whether these seemingly distinct operators exhibit comparable behavior in terms of Krylov complexity. The duality we focus on is the transverse field Ising model (TFIM) \cite{Pfeuty:1970qrn} and its dual \cite{Kitaev:2009war}, the Kitaev chain \cite{Kitaev:2000nmw}, a 1D system of free Majorana fermions. The Jordan-Wigner transformation \cite{Jordan:1928wi}, a well-known fermionization map, relates these two models by mapping the spin degrees of freedom in the TFIM to fermionic excitations in the Kitaev chain. In this mapping, local spin operators in the TFIM become highly non-local string operators in the fermionic theory, and vice versa. Despite this difference in locality, both theories share the same spectrum, raising a natural question: {\it do local operators in the Ising model and their non-local duals in the Kitaev chain exhibit similar Krylov complexity?}\\

\noindent
Naively, it would seem that answer to the above question is a simple yes.  Any observable may be expressed in terms of the eigenbasis for the respectively models and, since the energies are identical and the eigenstates may be mapped to each other exactly, one may expect the correlators to all match.  However, as highlighted in  \cite{Kitaev:2009war, Greiter2014}, though the models are mathematically equivalent they are physically distinct.  The distinction may be boiled down to the presence (absence) of topological order in the Kitaev chain (Ising model) and its robustness under local perturbations \cite{Kitaev:2009war}. With this additional knowledge, the question above becomes significantly more subtle.  As we will demonstrate, the naive intuition holds for the spread complexity of a time-evolved reference state chosen from one of the parity sectors, but not for the K-complexity of the associated operator in general.  \\ \\
The paper is organised as follows. In section 2 we introduce the Kitaev chain as well as the Ising chain and the dual Hamiltonian that it maps to under the Jordan-Wigner transformation.  We highlight the distinctions between the two as well as the operators that one may expect to probe this distinction.  In section 3 we briefly introduce our chosen diagnostic, Krylov complexity, and the methods we employ to compute it.  Section 4 contains the results of our investigation, and we finally conclude with discussion of these and possible future directions in section 5.

\section{Boson-Fermion Duality}

The concrete duality that we will focus on in this paper is the hardcore boson-fermion duality realised by the Jordan-Wigner transformation \cite{Jordan:1928wi}. The primary reason for this is that the systems related in this way are finite-dimensional so that the computation of complexity avoids some of the difficulties presented in bosonic systems \cite{Haque:2022ncl, Beetar:2023mfn}.  Furthermore, the specific pair of dual systems we have in mind, the Ising and Kitaev chains, have already been studied in the context of the complexity literature providing a nice way to benchmark our results \cite{Parker:2018yvk,Espanol:2022cqr,Bhattacharya:2023xjx,Caputa:2022yju,Zhang:2023wtr}.  In this paper, we will specifically be interested in understanding the behavior of K-complexity as viewed through the duality map. \\ \\
As articulated in \cite{Parker:2018yvk}, a duality transformation can map certain non-local operators on one side of the duality to local operators on the other side of the duality.  As such, the growth of operators may not be correlated with the growth of complexity.  Indeed, in integrable models the dynamics of operator growth may be restricted to a symmetry sector and operators classified according to the symmetry sector within which they fall.  An (imminently relevant) example of this may be found in the $L$-site Kitaev chain  with general twisted boundary conditions \cite{Kawabata:2017zsb}
\begin{eqnarray}
H_{K,b} & = &  \sum_{i=1}^{L-1}\left[   - t (c_{j+1}^\dag c_j + c_{j}^\dag c_{j+1} ) + \Delta (c_{j+1}^{\dag} c_j^\dag + c_{j} c_{j+1}) 
   \right] - \sum_{j=1}^L \mu (c_j^\dag c_j - c_j c_j^\dag)     \nonumber \\
& & + b\left[ -t (e^{i\phi_1} c_L^\dag c_1 + e^{-i\phi_1} c_1^\dag c_L ) + \Delta   (e^{i\phi_2} c_L^\dag c_1^\dag + e^{-i\phi_2} c_1 c_L )    \right]\;.  \label{HKitaevtwist} 
\end{eqnarray}
This Hamiltonian may be recast \cite{Kawabata:2017zsb} as a quadratic Hamiltonian in terms of the Majorana-like generators,
\begin{eqnarray}
\gamma^a_j & = & c_j + c_j^\dag    \nonumber \\
\gamma^b_j & = & i (c_j^\dag - c_j)\;.   \label{Majoranas}
\end{eqnarray}
Since it is quadratic, the dynamics of an operator of the form
\begin{equation}
\prod_{i=1}^{L} (\gamma^a_{i})^{n_i} (\gamma^b_{i})^{m_i}  \ \ \ ; \ \ \ n_i,m_i \in \left\{ 0,1 
 \right\}    \label{FermOps}
\end{equation}
is restricted to the subspace of fixed
\begin{equation}
n = \sum_{i} ( n_{i} + m_i ) \;,     \label{Fermspaces}
\end{equation}
which has dimension 
\begin{equation}
\textnormal{dim} =    \left( \begin{array}{c} 2L \\ n   \end{array}   \right) \;.  \label{FermDim}
\end{equation}
The interested reader may consult appendix \ref{MajoranaAppendix} for further details. 
A similar decomposition holds for states.  Since the Hamiltonian is quadratic in creation / annihilation operators it may be diagonalised by means of a Bogoliubov transformation.  The Fock space may be decomposed into subspaces consisting of $n$ Bogoliubov quasi-particles.  These subspaces do not mix under time-evolution.   In what follows when we make references to \textit{subspaces} we will refer to the subspace of operators unless stated otherwise.  This is important, since these spaces are rather different.  For example, the dimension of the space of operators is $4^L$ while the dimension of the space of states is $2^L$.  \\ \\
For $n < \frac{L}{2}$ the dimension of the subspace (\ref{FermDim}) is a good estimate for the number of operators explored during time-evolution, but it turns out be an overestimate otherwise.  In any event, the dynamics of operators in the Kitaev chain only explores a fraction of the full $4^{L}$-dimensional space of operators. This {\it Krylov subspace}, is bounded by this number and, as such, measures of complexity will inherit closely related bounds.  A duality transformation will typically reorganise the space of operators but keep their relations under multiplication (and commutation) in tact. Consequently, both the Kitaev chain (\ref{HKitaevtwist}) and its dual should exhibit dynamics restricted to subspaces of the same dimension.  In particular the largest Krylov subspace one may even hope to obtain has dimension bounded by $\frac{(2L)!}{(L!)^2}$.  \\ \\
Based on this, and the known duality between the Ising chain and Kitaev chain, one may expect the operator dynamics to be organisable in an identical way for the Ising chain, with each choice of operators confined to subspaces of the dimensions outlined above.  As we will demonstrate, this expectation is only partly true and can be spoilt due to boundary effects that become important when studying the operator dynamics.

\subsection{Jordan-Wigner transformation}

As mentioned, the dual systems we wish to study are related by the Jordan-Wigner transformation \cite{Jordan:1928wi,Mbeng:2020awt, nielsenFermionicCanonicalCommutation} which we now briefly summarise. Up to rotations in the spin matrices, the transformation is given by
\begin{equation}
c_j \;  = \; \frac{1}{2}\left( \prod_{i<j} \sigma_i^z  \right)(\sigma_j^x - i \sigma_j^y) \; \equiv \;  \frac{1}{2}\left( \prod_{i<j} \sigma_i^z  \right) \sigma^{-}_j  \,,   \label{JWString}
\end{equation}
with the inverse transformation given by 
\begin{eqnarray}
\sigma_j^z & = & c_j^\dag c_j -c_j c_j^\dag\,,    \nonumber \\
\sigma_j^x - i \sigma_j^y & = & \prod_{i < j} \left( c_i^\dag c_i - c_i c_i^\dag \right) c_j\,.
\end{eqnarray}
The operators above satisfy the standard commutation relations
\begin{eqnarray}
\left\{ c_i, c_j^\dag   \right\} & = & \delta_{ij}\,,    \nonumber \\
\left\{ c_i, c_j \right\} & = & 0\,,    \nonumber \\
\left[ \sigma_j^{\mu}, \sigma_k^\nu  \right] & = & 2 i \epsilon_{\mu\nu\rho} \delta_{jk} \sigma_k^\nu\,. \
\end{eqnarray}
The above mapping provides a way to map an arbitrary Hamiltonian constituted of fermions to a dual Hamiltonian of Pauli spin matrices (and vice versa).  The representation (\ref{JWString}) is called a Jordan-Wigner string and while it is a local operator in the fermionic representation, it is expressed as a non-local product of Pauli spin matrices.  Nearest-neighbour terms of $\sigma^x$ and $\sigma^y$ matrices, however, map neatly to nearest neighbour terms
\begin{equation}
\sigma_j^x \sigma_{j+1}^x = -(c_{j+1}^{\dag} c_j + c_{j}^{\dag} c_{j+1}) - (c_{j+1}^{\dag} c_j^\dag + c_{j} c_{j+1})\,.
\end{equation}
One special case, however, is the coupling between the last and first site on the chain, which maps as 
\begin{equation}
\sigma_L^x \sigma_1^x = \left( \prod_{i < L} (c_i^\dag c_i - c_i c_i^\dag )\right) (c_L^\dag + c_L)(c_1^\dag + c_1).  
\end{equation} 
This mapping of the boundary term will have a non-trivial impact on the operator growth, as we will demonstrate in due course.  

\subsection{Ising-Kitaev chain duality}

\label{dualSection}

The specific dual models that are the focus of this paper are the transverse field Ising chain \cite{Mbeng:2020awt},
\begin{equation}
H = J \sum_{i=1}^{L-1} \sigma_{j}^x \sigma_{j+1}^x + J g \sum_{i=1}^L \sigma_{j}^z + J a \sigma_{L}^x \sigma_{1}^x \label{HIsing}\;,
\end{equation}
which maps under the Jordan-Wigner transformation to the dual Hamiltonian
\begin{eqnarray}
H_{JW} & = &  J g \sum_{i=1}^L (c_j^\dag c_j - c_j c_j^\dag)    - J \sum_{i=1}^{L-1} (c_{j+1}^\dag c_j + c_{j}^\dag c_{j+1} ) - J \sum_{i=1}^{L-1} (c_{j+1}^{\dag} c_j^\dag + c_{j} c_{j+1})     \nonumber \\
& & + J a \left( \prod_{i =1}^L  (c_i^\dag c_i - c_i c_i^\dag )\right) ( c_{L} c^\dag_1 - c_1 c^\dag_L  + c_1 c_L - c_L^\dag c_1^\dag  )\,.\label{HKitaev} 
\end{eqnarray}
The parameter $a$ can be chosen to produce open, periodic or anti-periodic boundary conditions in the Ising chain for $a=0, 1, -1$ respectively.  It may in general take any real value if twisted boundary conditions are imposed.  Note that throughout we will be referring to the boundary conditions as periodic / open as they pertain to the Ising model.  Under the Jordan-Wigner transformation one thus obtains a Hamiltonian that is very close to the Kitaev chain (\ref{HKitaevtwist}).  When $a=0$ the open boundary Ising chain maps precisely to the open boundary Kitaev chain
\begin{equation}
\left. H_{JW}\right|_{a = 0} = H_{K,0}
\end{equation}
As is well-established \cite{Kitaev_2001, Kitaev:2009war} this model has a two-fold degeneracy dictated by a $Z_2$ symmetry.   The operators underpinning this \cite{Kitaev:2009war} on the Ising chain side is the spin flip operator $P_S = \prod_{i=1}^L \sigma_i^z$ and on the Kitaev side the parity operator $P_F = (-1)^{\sum_{i=1}^L c^\dag_i c_i}$.  \\ \\  
For $a \neq 0$ we also need to consider the boundary term.  The boundary term depends on the parity of the state on which it acts\footnote{To be explicit, the Fock space is the usual $\prod_{i=1}^L (c_i^\dag)^{n_i} |0\rangle$ with $n_j = 0,1$.  The even and odd subspace have $\sum_{i=1}^L n_i = 2k$ and $\sum_{i=1}^L n_i = 2k+1$ respectively}. Since the Hamiltonian does not mix different parity sectors, we may write
\begin{equation}
H_{JW} = H_{K,+a} + H_{K,-a} \ \ \ ; \ \ \ \left[ H_{K,+a} , H_{K,-a}  \right] = 0
\end{equation}
where $H_{K,+a}$ and $H_{K,-a}$ are obtained by projecting $H_{JW}$ onto the odd and even parity sectors respectively. One may replace the term $\prod_{i =1}^L  (c_i^\dag c_i - c_i c_i^\dag )$ by 1 or -1 when acting on states from the even or odd parity sector respectively.   Thus the dual Hamiltonian becomes precisely a Kitaev chain Hamiltonian when acting on one of these subspaces.  \\ \\
Consider now studying the return amplitude of a state chosen from one of the parity sectors.  To be explicit, consider the reference state
\begin{eqnarray}
|\psi\rangle & = & O |0\rangle    \nonumber \\
O & = & \prod_{i=1}^L (c_{i}^\dag)^{n_i}  \ \ \ ;   \ \ \ n_i = 0,1  \nonumber
\end{eqnarray}
The operator (and thus also the state) is from a definite parity sector since 
\begin{equation}
\left[ P_F,   O\right] = (-1)^{\sum_i n_i} O
\end{equation}
Consider now the return amplitude for this reference state.  This becomes
\begin{equation}
R(t) = \langle \psi | e^{-i t H_{JW} } |\psi \rangle  = \frac{1 - (-1)^{\sum_i n_i}}{2}\langle \psi | e^{-i t H_{K,+a}} |\psi\rangle + \frac{1 + (-1)^{\sum_i n_i}}{2}\langle \psi | e^{-i t H_{K,-a}} |\psi\rangle
\end{equation}
The implication of the above is that, for return amplitudes of this type, one may always replace the dynamics with that of the Kitaev chain with a fixed boundary condition\footnote{One may have considered a reference state that is a superposition of the even and odd parity sectors, in which case a simple replacement like this would not be possible and the spread complexity would behave differently.  However, our focus here is on the subtle difference between the state dynamics of $O(t) |0\rangle$  and the operator dynamics of $O(t)$ which one may expect (naively) to behave similarly.}.  \\ \\
For operators, however, the picture is quite different.  Consider the following expectation value of the operator $O$ w.r.t an arbitrary state from one of the parity sectors
\begin{align}
& \langle \psi_{+}| O e^{-i t H_{JW}} O e^{i t H_{JW}} |\psi_{+}\rangle  \nonumber   \\
=& \langle \psi_{+}| O e^{-i t H_{K,+a}} e^{-i t H_{K,-a}} O e^{i t H_{K,+a}} |\psi_{+}\rangle   \nonumber    \\
=& \sum_{|\phi_{+}\rangle} \langle \psi_{+}| O |\phi_{+}\rangle \langle \phi_{+} | e^{-i t H_{K,+a}} O e^{i t H_{K,+a}} |\psi_{+}\rangle + \sum_{|\phi_{-}\rangle} \langle \psi_{+}| O |\phi_{-}\rangle \langle \phi_{-} | e^{-i t H_{K,-a}} O e^{i t H_{K,+a}} |\psi_{+}\rangle
\label{operator inner product expansion}
\end{align}
 Importantly the replacement $H_{JW} \rightarrow H_{K,b}$ is now not always possible.  It is only correct if either the operator does not mix the different parity sectors i.e. $\langle \psi_{+}| O |\phi_{-}\rangle = 0$ or one is studying open boundary conditions. \\ \\
 This will have dramatic consequences.  When studying operators that mix states from the two parity sectors, we will find that the operators are no longer restricted to the subspace  described in (\ref{FermOps}),(\ref{Fermspaces}) and, due to the boundary term in (\ref{HKitaev}), can mix with other operators of the same parity.  This will give rise to operator complexity that is in contradiction with the intuition built up from the Kitaev chain. 
 To better understand the operator dynamics we will observe in the results section, it is useful to recast the Hamiltonian in terms of the operators (\ref{Majoranas}) which yields
 \begin{equation}
 H_{JW}  =  i Jg\sum_{i=1}^L \gamma^a_j \gamma^b_j + -i J\sum_{i=1}^{L-1} \gamma^a_j \gamma^b_{j+1} - i^{L+1} Ja(\prod_{j=1}^L \gamma^a_j \gamma^b_j) \gamma^a_L \gamma^b_{1}
 \end{equation}
 The Hamiltonian is thus constituted of a collection of quadratic terms connecting the bulk sites and a term consisting of $2L-2$ Majoranas.  This last term is the dual to the boundary term in the dual Ising chain.  \\ \\
 In what follows we will restrict to the critical Ising mode, $g=1$, which is sufficient to demonstrate the general effect when going from open to periodic boundary conditions.  The interested reader may consult \cite{PanditThesis2024} for details involving the strong and weak coupling regimes.  The conclusions we come to in this paper related to the subspaces explored are applicable away from the critical limit, though the detailed dynamics does depend on the values for the couplings.  

\section{Krylov Complexity}

Our chosen notion of complexity for this article is that of Krylov- or K-complexity \cite{Parker:2018yvk, Balasubramanian:2021mxo} which we now briefly summarise.  \\ \\
Consider some quantum Hamiltonian $H$ and a time-dependent Heisenberg operator $\OO(t)$ in the model. The evolution of the operator is given by the Heisenberg equations of motion
\begin{equation}
    \partial_t \OO(t)\,=\,i[H,\OO(t)]\,\,=i\LL \OO(t)\;,
    \label{eq:HeisenbergEoM}
\end{equation}
where $\LL=[H,\cdot]$ is the Liouvillian superoperator. The formal solution of this equation is then given by
\begin{align}
    \OO (t) &=e^{iHt} \OO(0) e^{-iHt} \nonumber \\
     &=e^{i\LL t} \OO \nonumber \\
     &=\sum^{\infty}_{n=0} \frac{(it)^n}{n!} \LL^{n} \OO\,.
    \label{eq:O(t)}
\end{align}
The operators $\LL^{n} \OO$ therefore provide a basis for the space of operators obtained by means of time-evolution and superposition.  We can naturally associate a vector space to these operators by making the association $\OO \rightarrow |\OO)$ and then specify an appropriate inner product to obtain a Hilbert space \cite{Caputa:2021sib}.  We will take this inner product to be the infinite temperature Wightman inner product, also known as the Frobenius inner product, $(A|B)=\text{Tr}(A^\dagger B)$.
The operators (\ref{eq:O(t)}) are not orthogonal with respect to this inner product and we may orthogonalise them by means of a Gram-Schmidt process to obtain the Krylov basis
\begin{equation}
|\OO_n) = \sum_{m=0}^n c_{m,n} |\LL^{n} \OO )\,; \ \ \ (\OO_m|\OO_n) = \delta_{m,n}\,.
\end{equation}
A noteworthy feature of the Krylov basis is that the Liouvillian takes on a tri-diagonal form \cite{Parker:2018yvk}
\begin{equation}
    \LL |\OO_n) = b_n|\OO_{n-1}) +b_{n+1} |\OO_{n+1})+ a_n |\OO_{n}) \; ,
    \label{eq:LL|OO_n)}
\end{equation}
so that the time-evolution of the operator $\OO (t)$ has been mapped to the evolution on a nearest-neighbour tight-binding spin chain.  The probability amplitudes $(\OO(t) | \OO_n)$ characterise the full operator evolution and the K-complexity is usually defined as the mean of this distribution
\begin{equation}
C(t; |\OO), \LL) = \sum_{n} n |(\OO(t) | \OO_n)|^2\,.    \label{KCompDef}
\end{equation}
Lastly, note that the probability amplitudes and Lanczos coefficients may be computed directly from the expectation value \cite{Muck:2022xfc}
\begin{equation}
R(t) = (\OO | \OO(t)).  
\end{equation}
In this way, many of the properties of K-complexity for different operators may be extracted by comparing the corresponding expectation values.  

\subsection{Symmetries}
\label{sec:Symmetries}

Our interest in this article will be on the case where the target operator is the time-evolved reference operator. Here, the K-complexity is, of course, a time-dependent quantity fully determined by the reference operator and Hamiltonian.  However, it is also known \cite{Sanchez-Garrido:2024pcy} that complexity is invariant under a simultaneous unitary transformation of the Hamiltonian and reference operator, 
\begin{equation}
C\left( t ; |\OO), \LL  \right) = C\left( t ; |U \OO U^\dag ), U \LL U^\dag  \right)\,.
\end{equation}
This implies that any transformation which commutes with the Hamiltonian furnishes a set of operators which provide identical expressions for K-complexity
\begin{equation}
C\left( t ; |U \OO U^\dag ), \LL  \right) = C\left( t ; |\OO), \LL  \right) \ \ \textnormal{if} \ \ \ [H, U] = 0    \label{equalForStates}
\end{equation}  
If a transformation commutes with both the Hamiltonian and the reference operator its dynamics is always restricted to the subspace of operators which commutes with this transformation i.e. 
\begin{equation}
\left[ U, \OO   \right] = 0 = \left[ U, H   \right] \ \ \ \Rightarrow \ \ \ \left[ U, e^{-i  \LL t } \OO  \right] = 0\,,     \nonumber
\end{equation}
which is inherited by the Krylov basis
\begin{equation}
\left[ U, \OO   \right] = 0 = \left[ U, H   \right] \ \ \ \Rightarrow \ \ \ \left[ U,  \OO_n  \right] = 0 \ \ \forall  \ n\,.
\end{equation}
A direct consequence of this is that the dimension of the Krylov subspace for such operators are lower than for a generic choice of initial operator.  \\ \\
Returning now to the model of interest for our purposes, the transverse field Ising model (\ref{HIsing}), the Hamiltonian is famously $\mathbb{Z}_{2}$-invariant under flipping the chain
\begin{equation}
H = \left. H \right|_{  {\sigma_i^\mu} \leftrightarrow \sigma_{L-i+1}^\mu  }
\label{eq:OBC flip symmetry}
\end{equation}
for any value of $a$.  For periodic boundary conditions, for which $a=1$, it also possesses a  translational symmetry 
\begin{equation}
H = \left. H \right|_{  {\sigma_i^\mu} \leftrightarrow \sigma_{i+1}^\mu  }
\label{eq:PBC translational symmetry}
\end{equation}
This, along with (\ref{equalForStates}) implies that the choice of reference operator is also invariant under these symmetries.  On the fermionic side of the duality things are more interesting.  Under reflection, 
\begin{equation}
H_{JW} = -\left. H_{JW} \right|_{  c_i \leftrightarrow c_{L-i+1}  }    \label{flipFermion}
\end{equation}
so that the Hamiltonian flips sign and amounts to time-reversal.  The presence of the boundary term, however, breaks translational symmetry in the strict sense.  Translational symmetry may be restored for certain choices of reference operator, depending on whether parity sectors are mixed or not.  

\subsection{Computing K-complexity}

Practically speaking, there are various ways in which K-complexity and spread complexity may be computed.  This is due to the fact that the Lanczos coefficients, return amplitude and Krylov probability amplitudes encode exactly the same information \cite{Parker:2018yvk}. Given any subset of these, various schemes can be designed to compute K-complexity \cite{Parker:2018yvk}. This is especially true in the case of finite-dimensional systems where the full set of Lanczos coefficients may be computed (at least in principle). One popular method is to make use of the Schrodinger-like equation,
\begin{equation}
  i \partial_t ( \OO_n | \OO(t) ) = a_n ( \OO_n | \OO_r(t) ) + b_n ( \OO_{n-1} | \OO_r(t) ) + b_{n+1}( \OO_{n+1} | \OO_r(t) )\,,   \label{SchrEqs}
\end{equation}
with boundary conditions $( \OO_0 | \OO(0) ) = 1$ and $( \OO_{N_{max}} | \OO(t) ) = 0 $, where $N_{max}$ is the dimension of the Krylov subspace.  From this set of equations the Krylov probability amplitudes can be computed and substituted into (\ref{KCompDef}) to obtain the K-complexity.  It is important to note that the equations (\ref{SchrEqs}) preserve probability, regardless of the values obtained for the Lanczos coefficients.  If an error (for example due to numerical precision loss or terminating the Lanczos algorithm prematurely) occurs, the probability amplitude will thus be perfectly well-behaved, though erroneous.  \\ \\
For this reason, we will primarily be computing the probability amplitudes using a different scheme\textemdash one that uses both the Lanczos coefficients and return amplitude, which we can compute numerically.  The probability amplitudes can be generated recursively through \cite{Beetar:2023mfn},
\begin{eqnarray}
 ( \OO_{n+1} | \OO_r(t) ) & = & \sum_{m=0}^{n+1} k_{m,n+1} \partial_t^m ( \OO_n | \OO_r(t) )  \nonumber \\
  k_{m,n+1} & = & \frac{i k_{m-1,n} - a_n k_{m,n} - b_{n} k_{m,n-1}}{b_{n+1}} \;.\label{LanczosProb}
\end{eqnarray}
A crucial feature of the this formula is that, since the return amplitude and Lanczos coefficients are directly related, any numerical error obtained in computing the Lanczos coefficients will be apparent.  Unlike the result from the Schr\"{o}dinger equation (\ref{SchrEqs}), the probability amplitudes only sum to 1 if all the Krylov probability amplitudes are computed in harmony with both these inputs.  Furthermore, one may terminate the set of Lanczos coefficients at a value smaller than the Krylov subspace dimension and an error may be estimated based on how close this smaller value of probability amplitudes are to $1$.  

\section{Results}   

\subsection{Notation}

Recall the operator $\sigma^-_j$ is the spin lowering operator in the Ising Chain acting on site j
\begin{equation}
    \sigma^-_j \; = \; \Bigg(\bigotimes_{l=1}^{j-1} I\Bigg) \otimes  \sigma^- \otimes \Bigg( \bigotimes_{l=j+1}^{L} I \Bigg) \;.
    \label{eqn: sigma_minus}
\end{equation}
And recall that the operator $\cdj$ is the fermionic annihilation operator in the dual Hamiltonian acting on site j, which we will represent as
\begin{equation}
    \cdj \; = \; \Bigg(\bigotimes_{l=1}^{j-1} \pz \Bigg) \otimes  \sigma^- \otimes \Bigg( \bigotimes_{l=j+1}^{L} I \Bigg) \;.
    \label{eqn: c_dagger_j}
\end{equation}

\noindent
We will denote the multi-site operators using a shorthand where the subscripts of an operator denote the location along the product chain at which the operator lies, and all other sites will contain the Identity. For instance,
\begin{equation}
    \sigma^+_{12} := \sigma^+ \otimes \sigma^+ \otimes I \otimes \ldots \otimes I \;.
    \label{eq:sp12}
\end{equation}

\noindent
We can compute the complexity of each of these operators, in the form above, with respect to the Ising Hamiltonian. Note that computing the complexity of a product of Jordan-Wigner strings with respect to the Ising Hamiltonian is exactly equivalent to computing the complexity of a product of $c_i^\dag$ operators with respect to $H_{JW}$ (\ref{HKitaev}).  We choose to use the Ising Hamiltonian since (in this Hilbert space of operators) all the operator matrices can be easily obtained via the usual tensor product.

\subsection{Open Boundary Condition}

\begin{figure}[h]
        \begin{center}
            \includegraphics[width=0.99\textwidth]{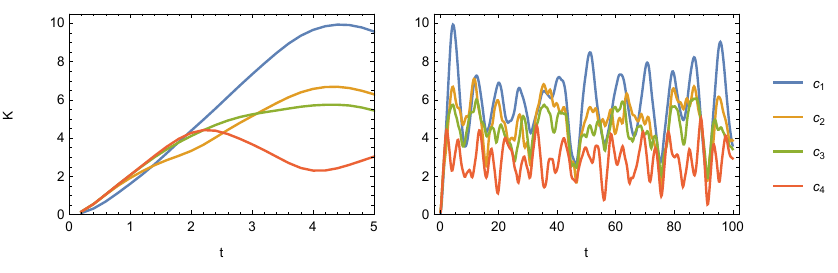}
        
        \caption{The early- and late-time K-Complexity of annihilation operators at various sites with respect to L=7 TFIM with OBC's.  The complexity is bounded by $2L = 14$ for all times.  The complexity of $c_4$ is further restricted to $L$ by reflection symmetry.  }
        \label{fig:ciShifted}
        \end{center}
\end{figure}

\begin{figure}[h]
    \centering
    \includegraphics[width=0.9\linewidth]{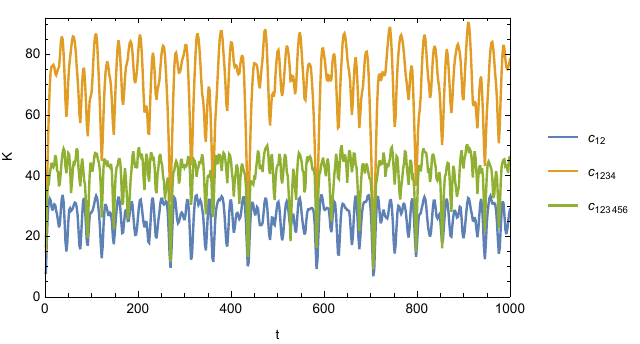}
    \caption{The Krylov complexity of a collection of even fermionic annihilation operators in the dual Hamiltonian with open boundary conditions and $L=7$.  The complexity increases with an increasing number of fermions in line with expectations\textemdash the size of the subspace of operators explored increases with the number of fermions up to $n = \frac{L+1}{2}$.}
    \label{fig:OBCfermionseven}
\end{figure}

\begin{figure}[h]
    \centering
    \includegraphics[width=0.9\linewidth]{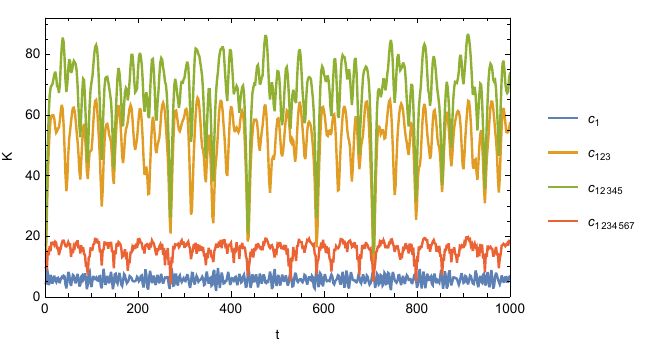}
    \caption{The Krylov complexity of a collection of odd fermionic annihilation operators in the dual Hamiltonian with open boundary conditions and $L=7$.  The complexity increases with an increasing number of fermions in line with expectations\textemdash the size of the subspace of operators explored increases with the number of fermions up to $n = \frac{L+1}{2}$.}
    \label{fig:OBCfermionsodd}
\end{figure}

We start our results in the open boundary condition setup.  A natural operator to study first is that of a single Jordan-Wigner string / fermionic operator.  In Fig. (\ref{fig:ciShifted}) we have plotted the Krylov complexity of a single fermionic annihilation operator at various sites for $L=7$.  
At very early times, only the complexity for fermions positioned next to the boundary is different and grows slightly more slowly. 
The complexity is bounded by $2L$ for all possible positions of the annihilation operator which is also the dimension of the subspace explored by the operator.  The complexity exhibits the $Z_2$ flip symmetry outlined in eq. (\ref{eq:OBC flip symmetry}), which explicitly implies identical expressions for the complexity of $c_i$ and $c_{L+1-i}$.  The complexity of $c_4$ (located at the centre of the $L=7$ chain) is bounded by $L$ which is the dimension of the subspace of operators invariant under this flip symmetry (\ref{flipFermion}).  Of course, with the open boundary conditions there is no expectation of translational symmetry for the operators.  \\ \\
Now let's consider what happens when we increase size of the reference operator.  Anticipating what we will find for the periodic case, we will discuss operators with an even number of fermions and an odd number of fermions separately.  In Figs. (\ref{fig:OBCfermionseven}) and (\ref{fig:OBCfermionsodd})  we have plotted the K-complexity of a collection of fermions for an $L=7$ lattice.  These results are in complete harmony\footnote{As mentioned, the complexity is bounded by the dimension of these subspaces.  For the small reference operators the dimension of the Krylov space is close to this bound, while larger reference operators explore a more restricted subset of the subspace operators.} with the restriction to subspaces discussed around eq. (\ref{Fermspaces}).  As the number of fermions are increased, the size of the subspace it can explore is increased, which often results in a larger K-complexity saturation value.  The complexity of an operator can never exceed the dimension of the subspace to which it is confined.  Of course, it may only explore a smaller subset of operators, as is the case for the products of fermions that are larger than half the length of the chain.  \\ \\
In summary, the complexity for an arbitrary operator in an Ising chain with open boundary conditions behaves precisely as one would expect for an integrable system: the operator dynamics restricted to subspaces that may be characterised cleanly in terms of the dual fermionic description.  The dimension of these subspaces provide clear upper bounds for complexity.  We note that, on the Ising side of the duality the operators are, of course, organised according to these same subspaces though the subspace is no longer correlated with the number of occupied sites. Rather, the operator space is organised by the number of $\sigma_x$, $\sigma_y$ operators and twice the number of "misplaced" $\sigma_0$ and $\sigma_z$ operators, see appendix \ref{PauliAppendix} for details.

\subsection{Periodic Boundary conditions}

\begin{figure}[h]
        \begin{center}
            \includegraphics[width=0.99\textwidth]{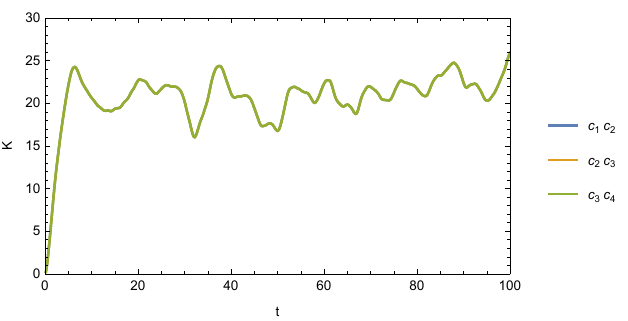}
        \end{center}
        \caption{The K-Complexity for a pair of adjacent creation operators at various sites with respect to L=7 TFIM with PBCs. The translational symmetry stems from the fact that the operators have even fermionic parity, and thus do not mix parity sectors.}
        \label{L7pairsPeriodic}
\end{figure}

\begin{figure}[h]
        \begin{center}
            \includegraphics[width=0.8\textwidth]{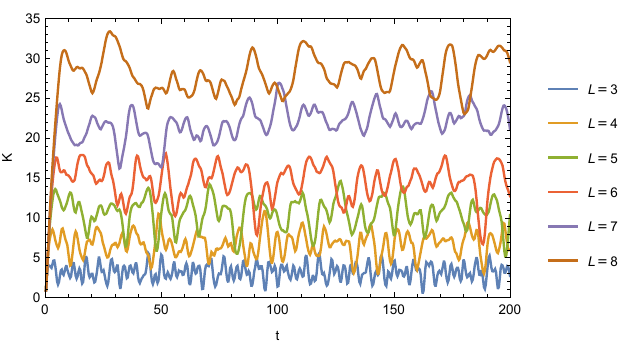}
        \end{center}
        \caption{The K-Complexity for $c_1(t) c_2(t)$ for various values of $L$ and periodic boundary conditions. The growth with increasing $L$ is predicted to be a polynomial in $L$, giving rise to rather modest growth.}
        \label{c1c2forDifferentL}
\end{figure}

\begin{figure}[h]
        \begin{center}
            \includegraphics[width=0.99\textwidth]{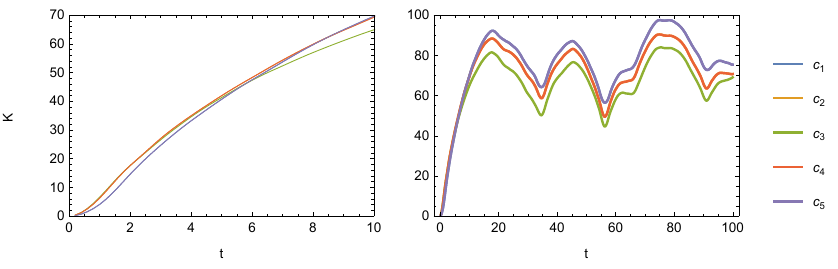}
        \end{center}
        \caption{The early- and late-time K-Complexity of the creation operators at various sites with respect to L=5 TFIM with PBCs. K-Complexity of $c_1$ is exactly equivalent to all $\sigma^{-}_j$ due to the translational symmetry.  The complexity of $c_j$ for different $j$ are, however, not equal and translational symmetry on the fermion side is broken.}
        \label{fig:jwL5periodic}
\end{figure}

\begin{figure}[h]
        \begin{center}
            \includegraphics[width=0.8\textwidth]{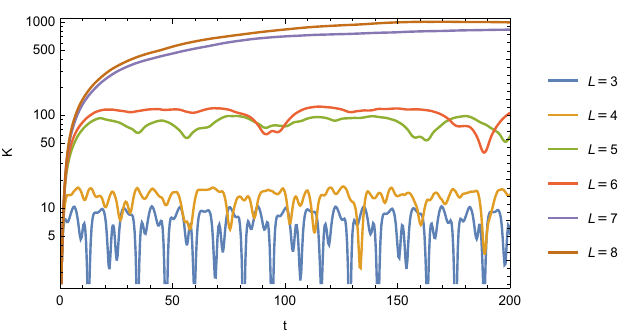}
        \end{center}
        \caption{The K-Complexity $c_1(t)$ for various values of $L$ and periodic boundary conditions.  As $L$ increases the value of complexity increases dramatically.  For every odd $L$ the operator dynamics involve additional subspaces while the increase form odd to even $L$ is only due to the increase in size of these subspaces.}
        \label{c1forDifferentL}
\end{figure}

We now turn our attention to the periodic Ising chain.  As highlighted in section \ref{dualSection}, the boundary condition obtained under the duality transformation will give rise to non-trivial effects when considering operators that mix parity sectors \footnotemark.  First, we confirm that things behave more or less as expected for operators that do not mix parity sectors.  In Fig. (\ref{L7pairsPeriodic}) we have plotted the K-complexity for pairs of adjacent fermionic operators\textemdash these have even fermion parity, and will thus not mix parity sectors.  \footnotetext{Operators which mix parity sectors refer to operators that utilize both parity sectors of the Hamiltonian in the inner product, as seen in eq. (\ref{operator inner product expansion}).}
Two features are noteworthy: first, these operators respect periodic / anti-periodic symmetry similar to their Ising chain counterparts.  We anticipated this in the discussion around eq. (\ref{operator inner product expansion}).  Furthermore, the saturation value of complexity is bounded by $\frac{2L(2L-1)}{2}$ being the dimension of the subspace of fermionic operators they are confined to.  In Fig. (\ref{c1c2forDifferentL}) we have plotted the complexity for various values of $L$ demonstrating this modest growth with chain length. 
\\ \\
In stark contrast, neither of these properties are true for parity odd operators such as single fermion operators.  In Fig. (\ref{fig:jwL5periodic}) we have plotted the complexity for single fermions at various sites in an $L=5$ chain. The complexity demonstrates the reflection symmetry highlighted in (\ref{flipFermion}) but no periodic symmetry.  As discussed, this is due to the single fermionic operators mixing the two parity sectors.  Secondly, the operator is no longer confined to the subspaces described in (\ref{Fermspaces}) and can, in principle, explore the full space of parity-mixing operators.  This is gives rise to an enormous increase in the dimension of the Krylov subspace and, subsequently, the saturation value of Krylov complexity.  For comparison, the corresponding complexity in the open boundary condition is an order of magnitude less for open boundary conditions, see Fig (\ref{fig:ciShifted}).
At large $L$ this subspace constitutes about half of the total operators and thus grows as $2^L$ at large $L$.  Due to finite computational power we have only plotted this for values of $L$ far from this limit in Fig. (\ref{c1forDifferentL}), but the dramatic increase in operator complexity is apparent with increasing $L$.  The increase in complexity is most stark between even length chains $L=2k$ and the next integer, $L=2k+1$.  In this case the number of subspaces for parity-mixing operators is increased by one.  When going from $L=2k+1$ to $L=2k+2$ the number of parity-mixing subspaces remains the same, though the size of each subspace has grown.  This gives rise only to a modest increase in complexity. \\ \\

\begin{figure}[h]
        \begin{center}
            \includegraphics[width=0.99\textwidth]{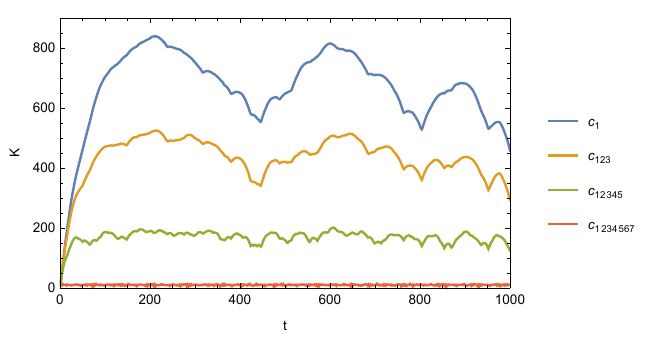}
                        \caption{The K-Complexity for a sequence of Jordan-Wigner strings for an $L=7$ periodic Ising chain.  With a reference operator consisting of an odd number of JW strings the parity sectors mix under the boundary term  }
        \label{L7Oddplot}
        \end{center}
\end{figure}

\noindent
When we start increasing the size of the initial operator we note an interesting feature: the complexity of the single Jordan-Wigner string is larger than any other sequence, see Fig. (\ref{L7Oddplot}).  This is the exact opposite of the case for the open boundary condition, Fig. (\ref{fig:OBCfermionsodd}), where the complexity of the single JW string was the smallest, since it was confined to the smallest subspace.  In the case of periodic boundary conditions, it can grow to access the other subspaces.  To understand why it is larger than any other (odd-numbered) collection of fermions, the results in appendix \ref{MajoranaAppendix} is useful, specifically eq. (\ref{MajoranaCommsboundary}).  The boundary term is responsible for the operator growth of odd-parity operators.  More precisely, though, odd-parity operators that involve both sites $1$ and $L$ and those that involve neither do not mix under commutations with the boundary term.  For a single fermion to grow to access all the odd-operator subspaces it requires a sequence of specific commutators.  By the commutator with the boundary term it can immediately access the subspace with $2L-3$ Majoranas, but is still restricted to a smaller subspace of operators not involving sites $1$ and $L$.  Commutators with the hopping terms now allow this operator to wrap around the boundary, after which a commutator with the boundary term yields an operator in the subspace of $3$ Majoranas which does involve sites $1$ and $L$.  After the hopping terms unwrap it, one can then access the subspace of $2L-5$ operators by a commutator with the boundary term and so on.
 This sequence of commutators mean that the largest subspaces ($n\sim L$) are the ones that are captured by the Krylov basis last and thus they are more complex.  For larger starting operators these large subspaces are accessed earlier in the Lanczos algorithm, and are thus less complex.  
\begin{figure}[h]
    \begin{center}
            \includegraphics[width=0.99\textwidth]{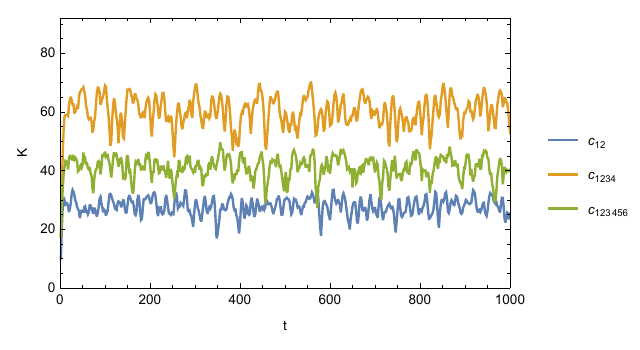}
    \end{center}
            \caption{The K-Complexity for a sequence of Jordan-Wigner strings for an $L=7$ periodic Ising chain.  An even number of strings does not mix parity sectors and thus the complexity is small, bounded by a polynomial in $L$. }
        \label{L7evenplot}
\end{figure}
\\ \\
We emphasise that the mechanism for operator growth of the single fermion is due, entirely, to the effect of the boundary term under the duality transformation.  Consider a Jordan-Wigner string prepared in a large $L$ periodic transverse field Ising model with a "tail" of many $\sigma_z$ operators.  This operator evolves, initially, like a single fermionic operator in the open boundary condition setup with a tight bound on its complexity.  As the overlap with the boundary grows it can, however, access many higher-complexity states and grow into an operator built as the product of many Jordan-Wigner strings.  The interplay between operator growth and the nature of the boundary condition may have some application in setups where the boundary can be tuned between open and periodic in a dynamic way.  Note also that this effect will be present, independent of the strength of the connection between site $1$ and $L$.

\section{Conclusions}
In this article, we have explored the growth of Krylov complexity of operators in dual theories, focusing on the transverse field Ising model and its fermionic dual, the Kitaev chain. Our study probed the hypothesis that the non-local operators which are dual to local operators necessarily display complexity attributes similar to that of local operators. In the case of open boundary conditions our finding is that this statement is true. The duality map re-organizes the space of operators, but for each (local) operator on the one side one can find a dual operator with identical K-complexity in the Ising chain Hamiltonian. The re-organization of the Hilbert space results in operators which we consider to be simple, local operators of each dual theory displaying vastly different complexity behaviour. In our example the growth of any choice of operator is tightly constrained by the quadratic nature of the Kitaev chain Hamiltonian. Specifically, the dynamics of an operator is constrained to some subspace occupying only a small fraction of the space of operators.\\

\noindent
The dual of the transverse field Ising model with periodic boundary conditions is not precisely the Kitaev chain, since the boundary term depends on the parity sector on which it acts. Operators which mix parity sectors are now enabled to access many branches of operators, in stark contrast to its behaviour with open boundary conditions. This in turn leads to a significant increase in the saturation values of complexity. This observation suggests that boundary effects and the structure of dual operators have profound implications for the late-time behaviour of complexity, further emphasizing the rich interplay between locality, duality, and operator growth.\\

\noindent
Our results provide new insights into how duality affects the dynamics of operator growth in quantum systems. While dual theories are expected to share similar spectra, our work shows that operator growth, as quantified by Krylov complexity, retains subtle distinctions that emerge due to the locality of operators and boundary effects. These findings not only deepen our understanding of Krylov complexity in dual theories but also suggest broader applications in the study of integrable systems and quantum chaos. Future work could explore these ideas in more complex dualities, such as those found in higher-dimensional systems and holography, where similar questions of operator growth and complexity are expected to play a key role.

\acknowledgments

We would like to thank Cameron Beetar and Jordan Cotler for insightful discussions.  JM and HJRVZ are supported in part by the ``Quantum Technologies for Sustainable Development''  grant from the National Institute for Theoretical and Computational Sciences of South Africa (NITHECS).

\bibliographystyle{JHEP}
\bibliography{complexityRefs}

\providecommand{\href}[2]{#2}\begingroup\raggedright\begin{thebibliography}{10}

\bibitem{Maldacena:2015waa}
J.~Maldacena, S.H.~Shenker and D.~Stanford, \emph{{A bound on chaos}},
  \href{https://doi.org/10.1007/JHEP08(2016)106}{\emph{JHEP} {\bfseries 08}
  (2016) 106} [\href{https://arxiv.org/abs/1503.01409}{{\ttfamily
  1503.01409}}].

\bibitem{Hashimoto:2017oit}
K.~Hashimoto, K.~Murata and R.~Yoshii, \emph{{Out-of-time-order correlators in
  quantum mechanics}},
  \href{https://doi.org/10.1007/JHEP10(2017)138}{\emph{JHEP} {\bfseries 10}
  (2017) 138} [\href{https://arxiv.org/abs/1703.09435}{{\ttfamily
  1703.09435}}].

\bibitem{Roberts:2018mnp}
D.A.~Roberts, D.~Stanford and A.~Streicher, \emph{{Operator growth in the SYK
  model}}, \href{https://doi.org/10.1007/JHEP06(2018)122}{\emph{JHEP}
  {\bfseries 06} (2018) 122}
  [\href{https://arxiv.org/abs/1802.02633}{{\ttfamily 1802.02633}}].

\bibitem{Parker:2018yvk}
D.E.~Parker, X.~Cao, A.~Avdoshkin, T.~Scaffidi and E.~Altman, \emph{{A
  Universal Operator Growth Hypothesis}},
  \href{https://doi.org/10.1103/PhysRevX.9.041017}{\emph{Phys. Rev. X}
  {\bfseries 9} (2019) 041017}
  [\href{https://arxiv.org/abs/1812.08657}{{\ttfamily 1812.08657}}].

\bibitem{Barbon:2019wsy}
J.L.F.~Barb\'on, E.~Rabinovici, R.~Shir and R.~Sinha, \emph{{On The Evolution
  Of Operator Complexity Beyond Scrambling}},
  \href{https://doi.org/10.1007/JHEP10(2019)264}{\emph{JHEP} {\bfseries 10}
  (2019) 264} [\href{https://arxiv.org/abs/1907.05393}{{\ttfamily
  1907.05393}}].

\bibitem{Dymarsky:2019elm}
A.~Dymarsky and A.~Gorsky, \emph{{Quantum chaos as delocalization in Krylov
  space}}, \href{https://doi.org/10.1103/PhysRevB.102.085137}{\emph{Phys. Rev.
  B} {\bfseries 102} (2020) 085137}
  [\href{https://arxiv.org/abs/1912.12227}{{\ttfamily 1912.12227}}].

\bibitem{PhysRevE.104.034112}
J.D.~Noh, \emph{Operator growth in the transverse-field ising spin chain with
  integrability-breaking longitudinal field},
  \href{https://doi.org/10.1103/PhysRevE.104.034112}{\emph{Phys. Rev. E}
  {\bfseries 104} (2021) 034112}.

\bibitem{Rabinovici:2020ryf}
E.~Rabinovici, A.~S\'anchez-Garrido, R.~Shir and J.~Sonner, \emph{{Operator
  complexity: a journey to the edge of Krylov space}},
  \href{https://doi.org/10.1007/JHEP06(2021)062}{\emph{JHEP} {\bfseries 06}
  (2021) 062} [\href{https://arxiv.org/abs/2009.01862}{{\ttfamily
  2009.01862}}].

\bibitem{Caputa:2021sib}
P.~Caputa, J.M.~Magan and D.~Patramanis, \emph{{Geometry of Krylov
  complexity}},
  \href{https://doi.org/10.1103/PhysRevResearch.4.013041}{\emph{Phys. Rev.
  Res.} {\bfseries 4} (2022) 013041}
  [\href{https://arxiv.org/abs/2109.03824}{{\ttfamily 2109.03824}}].

\bibitem{Balasubramanian:2022tpr}
V.~Balasubramanian, P.~Caputa, J.~Magan and Q.~Wu, \emph{{Quantum chaos and the
  complexity of spread of states}},
  \href{https://arxiv.org/abs/2202.06957}{{\ttfamily 2202.06957}}.

\bibitem{Nandy:2024htc}
P.~Nandy, A.S.~Matsoukas-Roubeas, P.~Mart\'\i{}nez-Azcona, A.~Dymarsky and
  A.~del Campo, \emph{{Quantum Dynamics in Krylov Space: Methods and
  Applications}},  \href{https://arxiv.org/abs/2405.09628}{{\ttfamily
  2405.09628}}.

\bibitem{viswanath2008recursion}
V.~Viswanath and G.~M{\"u}ller, \emph{The recursion method: application to
  many-body dynamics}, vol.~23, Springer Science \& Business Media (2008).

\bibitem{Ali:2019zcj}
T.~Ali, A.~Bhattacharyya, S.S.~Haque, E.H.~Kim, N.~Moynihan and J.~Murugan,
  \emph{{Chaos and Complexity in Quantum Mechanics}},
  \href{https://doi.org/10.1103/PhysRevD.101.026021}{\emph{Phys. Rev. D}
  {\bfseries 101} (2020) 026021}
  [\href{https://arxiv.org/abs/1905.13534}{{\ttfamily 1905.13534}}].

\bibitem{Bhattacharyya:2020kgu}
A.~Bhattacharyya, S.~Das, S.S.~Haque and B.~Underwood, \emph{{Rise of
  cosmological complexity: Saturation of growth and chaos}},
  \href{https://doi.org/10.1103/PhysRevResearch.2.033273}{\emph{Phys. Rev.
  Res.} {\bfseries 2} (2020) 033273}
  [\href{https://arxiv.org/abs/2005.10854}{{\ttfamily 2005.10854}}].

\bibitem{Bhattacharyya:2020rpy}
A.~Bhattacharyya, S.~Das, S.~Shajidul~Haque and B.~Underwood,
  \emph{{Cosmological Complexity}},
  \href{https://doi.org/10.1103/PhysRevD.101.106020}{\emph{Phys. Rev. D}
  {\bfseries 101} (2020) 106020}
  [\href{https://arxiv.org/abs/2001.08664}{{\ttfamily 2001.08664}}].

\bibitem{Bhattacharyya:2020art}
A.~Bhattacharyya, W.~Chemissany, S.S.~Haque, J.~Murugan and B.~Yan, \emph{{The
  Multi-faceted Inverted Harmonic Oscillator: Chaos and Complexity}},
  \href{https://doi.org/10.21468/SciPostPhysCore.4.1.002}{\emph{SciPost Phys.
  Core} {\bfseries 4} (2021) 002}
  [\href{https://arxiv.org/abs/2007.01232}{{\ttfamily 2007.01232}}].

\bibitem{Bhattacharyya:2020qtd}
A.~Bhattacharyya, S.S.~Haque, G.~Jafari, J.~Murugan and D.~Rapotu,
  \emph{{Krylov complexity and spectral form factor for noisy random matrix
  models}}, \href{https://doi.org/10.1007/JHEP10(2023)157}{\emph{JHEP}
  {\bfseries 23} (2020) 157}
  [\href{https://arxiv.org/abs/2307.15495}{{\ttfamily 2307.15495}}].

\bibitem{Bhattacharjee:2022vlt}
B.~Bhattacharjee, X.~Cao, P.~Nandy and T.~Pathak, \emph{{Krylov complexity in
  saddle-dominated scrambling}},
  \href{https://doi.org/10.1007/JHEP05(2022)174}{\emph{JHEP} {\bfseries 05}
  (2022) 174} [\href{https://arxiv.org/abs/2203.03534}{{\ttfamily
  2203.03534}}].

\bibitem{Bhattacharya:2023zqt}
A.~Bhattacharya, P.~Nandy, P.P.~Nath and H.~Sahu, \emph{{On Krylov complexity
  in open systems: an approach via bi-Lanczos algorithm}},
  \href{https://doi.org/10.1007/JHEP12(2023)066}{\emph{JHEP} {\bfseries 12}
  (2023) 066} [\href{https://arxiv.org/abs/2303.04175}{{\ttfamily
  2303.04175}}].

\bibitem{Sanchez-Garrido:2024pcy}
A.~S\'anchez-Garrido, \emph{{On Krylov Complexity}},  other thesis, 7, 2024,
  [\href{https://arxiv.org/abs/2407.03866}{{\ttfamily 2407.03866}}].

\bibitem{Avdoshkin:2022xuw}
A.~Avdoshkin, A.~Dymarsky and M.~Smolkin, \emph{{Krylov complexity in quantum
  field theory, and beyond}},
  \href{https://doi.org/10.1007/JHEP06(2024)066}{\emph{JHEP} {\bfseries 06}
  (2024) 066} [\href{https://arxiv.org/abs/2212.14429}{{\ttfamily
  2212.14429}}].

\bibitem{Pfeuty:1970qrn}
P.~Pfeuty, \emph{{The one-dimensional Ising model with a transverse field}},
  \href{https://doi.org/10.1016/0003-4916(70)90270-8}{\emph{Annals Phys.}
  {\bfseries 57} (1970) 79}.

\bibitem{Kitaev:2009war}
A.~Kitaev and C.~Laumann, \emph{{Topological phases and quantum computation}},
  4, 2009 [\href{https://arxiv.org/abs/0904.2771}{{\ttfamily 0904.2771}}].

\bibitem{Kitaev:2000nmw}
A.~Kitaev, \emph{{Unpaired Majorana fermions in quantum wires}},
  \href{https://doi.org/10.1070/1063-7869/44/10S/S29}{\emph{Phys. Usp.}
  {\bfseries 44} (2001) 131}
  [\href{https://arxiv.org/abs/cond-mat/0010440}{{\ttfamily
  cond-mat/0010440}}].

\bibitem{Jordan:1928wi}
P.~Jordan and E.P.~Wigner, \emph{{About the Pauli exclusion principle}},
  \href{https://doi.org/10.1007/BF01331938}{\emph{Z. Phys.} {\bfseries 47}
  (1928) 631}.

\bibitem{Greiter2014}
M.~Greiter, V.~Schnells and R.~Thomale, \emph{The 1d ising model and
  topological order in the kitaev chain},
  \href{https://doi.org/10.1016/j.aop.2014.08.013}{\emph{Annals of Physics}
  {\bfseries 351} (2014) }.

\bibitem{Haque:2022ncl}
S.S.~Haque, J.~Murugan, M.~Tladi and H.J.R.~Van~Zyl, \emph{{Krylov Complexity
  for Jacobi Coherent States}},
  \href{https://arxiv.org/abs/2212.13758}{{\ttfamily 2212.13758}}.

\bibitem{Beetar:2023mfn}
C.~Beetar, N.~Gupta, S.S.~Haque, J.~Murugan and H.J.R.~Van~Zyl,
  \emph{{Complexity and Operator Growth for Quantum Systems in Dynamic
  Equilibrium}},  \href{https://arxiv.org/abs/2312.15790}{{\ttfamily
  2312.15790}}.

\bibitem{Espanol:2022cqr}
B.L.~Espa\~nol and D.A.~Wisniacki, \emph{{Assessing the saturation of Krylov
  complexity as a measure of chaos}},
  \href{https://doi.org/10.1103/PhysRevE.107.024217}{\emph{Phys. Rev. E}
  {\bfseries 107} (2023) 024217}
  [\href{https://arxiv.org/abs/2212.06619}{{\ttfamily 2212.06619}}].

\bibitem{Bhattacharya:2023xjx}
A.~Bhattacharya, P.P.~Nath and H.~Sahu, \emph{{Krylov complexity for nonlocal
  spin chains}}, \href{https://doi.org/10.1103/PhysRevD.109.066010}{\emph{Phys.
  Rev. D} {\bfseries 109} (2024) 066010}
  [\href{https://arxiv.org/abs/2312.11677}{{\ttfamily 2312.11677}}].

\bibitem{Caputa:2022yju}
P.~Caputa, N.~Gupta, S.S.~Haque, S.~Liu, J.~Murugan and H.J.R.~Van~Zyl,
  \emph{{Spread Complexity and Topological Transitions in the Kitaev Chain}},
  \href{https://arxiv.org/abs/2208.06311}{{\ttfamily 2208.06311}}.

\bibitem{Zhang:2023wtr}
R.~Zhang and H.~Zhai, \emph{{Universal hypothesis of autocorrelation function
  from Krylov complexity}},
  \href{https://doi.org/10.1007/s44214-024-00054-4}{\emph{Quant. Front.}
  {\bfseries 3} (2024) 7} [\href{https://arxiv.org/abs/2305.02356}{{\ttfamily
  2305.02356}}].

\bibitem{Kawabata:2017zsb}
K.~Kawabata, R.~Kobayashi, N.~Wu and H.~Katsura, \emph{{Exact zero modes in
  twisted Kitaev chains}},
  \href{https://doi.org/10.1103/PhysRevB.95.195140}{\emph{Phys. Rev. B}
  {\bfseries 95} (2017) 195140}
  [\href{https://arxiv.org/abs/1702.00197}{{\ttfamily 1702.00197}}].

\bibitem{Mbeng:2020awt}
G.B.~Mbeng, A.~Russomanno and G.E.~Santoro, \emph{{The quantum Ising chain for
  beginners}},
  \href{https://doi.org/10.21468/SciPostPhysLectNotes.82}{\emph{SciPost Phys.
  Lect. Notes} {\bfseries 82} (2024) 1}
  [\href{https://arxiv.org/abs/2009.09208}{{\ttfamily 2009.09208}}].

\bibitem{nielsenFermionicCanonicalCommutation}
M.A.~Nielsen, \emph{The {{Fermionic}} canonical commutation relations and the
  {{Jordan-Wigner}} transform}, .

\bibitem{Kitaev_2001}
A.Y.~Kitaev, \emph{Unpaired majorana fermions in quantum wires},
  \href{https://doi.org/10.1070/1063-7869/44/10s/s29}{\emph{Physics-Uspekhi}
  {\bfseries 44} (2001) 131}.

\bibitem{PanditThesis2024}
Z.~Pandit, \emph{K-complexity and the jordan-wigner transformation},  Master's
  thesis, University of Cape Town, 2024.

\bibitem{Balasubramanian:2021mxo}
V.~Balasubramanian, M.~DeCross, A.~Kar, Y.C.~Li and O.~Parrikar,
  \emph{{Complexity growth in integrable and chaotic models}},
  \href{https://doi.org/10.1007/JHEP07(2021)011}{\emph{JHEP} {\bfseries 07}
  (2021) 011} [\href{https://arxiv.org/abs/2101.02209}{{\ttfamily
  2101.02209}}].

\bibitem{Muck:2022xfc}
W.~M\"uck and Y.~Yang, \emph{{Krylov complexity and orthogonal polynomials}},
  \href{https://doi.org/10.1016/j.nuclphysb.2022.115948}{\emph{Nucl. Phys. B}
  {\bfseries 984} (2022) 115948}
  [\href{https://arxiv.org/abs/2205.12815}{{\ttfamily 2205.12815}}].

\end{thebibliography}\endgroup


\appendix

\section{Majorana commutators}

\label{MajoranaAppendix}

Here we would like to state, for the interested reader, a useful identity when computing the commutator involving sequences of Majorana fermions.  Consider the shorthand
\begin{equation}
\gamma^I = (\gamma_1)^{i_1} (\gamma_2)^{i_2} (\gamma_3)^{i_3} \cdots (\gamma_n)^{i_n} \ \ \ I = \left\{i_1, i_2, \cdots, i_n   \right\}
\end{equation}
Two such terms satisfy the following identity
\begin{eqnarray}
\gamma^{I_1} \gamma^{I_2} & = & (-1)^\sigma \gamma^{J_1} \gamma^K \gamma^{K} \gamma^{J_2}    \ \ \ \ \ \ \textnormal{where} \ J_1 \cap J_2 = 0 \nonumber \\
& = & (-1)^\sigma (-1)^{|K|(|J_1| + |J_2|) }\gamma^K \gamma^{J_1} \gamma^{J_2} \gamma^K   \nonumber \\
& = & (-1)^\sigma (-1)^{|K|(|J_1| + |J_2|) } (-1)^{|J_1||J_2|} \gamma^K  \gamma^{J_2} \gamma^{J_1} \gamma^K \nonumber \\
& = &  (-1)^{ |K|(|J_1| + |J_2|)  } (-1)^{|J_1||J_2|}\gamma^{I_2} \gamma^{I_1}    \nonumber \\
&=&  (-1)^{ |I_1 \cap I_2|(|I_1| + |I_2| - 2 |I_1 \cap I_2| )  } (-1)^{(|I_1| - |I_1 \cap I_2| )(|I_2| - |I_1 \cap I_2| )}\gamma^{I_2} \gamma^{I_1}    \nonumber \\
& = &  (-1)^{|I_1||I_2|}(-1)^{|I_1 \cap I_2|}  \gamma^{I_2} \gamma^{I_1}
\end{eqnarray}
so that
\begin{equation}
\left[ \gamma^{I_1}, \gamma^{I_2}  \right] = ((-1)^{|I_1||I_2|}(-1)^{|I_1 \cap I_2|} - 1) \gamma^{I_2 \cup I_1 - I_2 \cap I_1} 
\end{equation}
Two examples of the above are particularly relevant for understanding the operator dynamics in the main text.  For the first , consider the case where $I_2$ consist of two Majorana fermions.  The above implies that this will commute with $\gamma^{I_1}$ unless there is exactly one element in $I_2 \cap I_1$.  For this case we also have that $|I_2 \cup I_1 - I_2 \cap I_1| = |I_1|$ so that the number of Majoranas in $I_1$ is conserved under this commutator.  \\ \\
Second, consider the full set of Majoranas $\gamma^J = \gamma^1 \gamma^2 \gamma^3 \cdots \gamma^n$ with $n = 2k$ even.  Using he identity above it is clear that any set with an even number of Majoranas commutes with $\gamma^J$. 
 Consider $\gamma^{I_1}$ to be a set with an odd number of Majoranas.  One can now show that 
 \begin{eqnarray}
 \left[ \gamma^{I_1}, \gamma^I \gamma^{I_2}  \right] & = & -(1 + (-1)^{|I_1 \cap I_2}|)\gamma^I \gamma^{I_1} \gamma^{I_2}    \nonumber \\
 & \sim & (1 + (-1)^{|I_1 \cap I_2}|) \gamma^{I - I_1 - I_2 + I_1 \cap I_2}   \nonumber
 \end{eqnarray}
 Now consider the case where $I_2 = \gamma^1 \gamma^{2k}$.  In this case we thus find
 \begin{eqnarray}
 \left[ \gamma^{I_1}, \gamma^I \gamma^{1} \gamma^{2k}  \right] & \sim &  \gamma^{I / \left\{1, 2k \right\} - I_1}     \ \ \ I_1 \cap \left\{1, 2k \right\}  = 0    \nonumber \\
 & \sim &  \gamma^{I - I_1 / \left\{1, 2k \right\}}     \ \ \ I_1 \cap \left\{1, 2k \right\}  = \left\{1, 2k \right\}    \label{MajoranaCommsboundary}
 \end{eqnarray}
 so that sequences that contain the "boundary" Majoranas $\gamma^1, \gamma^{2k}$ will continue to contain them under this commutator and ones that do not will continue not to contain them.

\section{Spin Chain Operator subspaces}

\label{PauliAppendix}

As shown in the main text, the organisation of subspaces is most clear in terms of the Majorana fermions, see eq. (\ref{Fermspaces}).  The corresponding subspace on the Ising chain side of the duality is obtained, of course, by performing the inverse Jordan-Wigner transformation.  Let's first consider the first few examples before stating the general pattern.  The $n=1$ subspace maps to
\begin{equation}
\left(\prod_{i=1}^{j-1} \sigma_i^z \right) \sigma^a_j \left(\prod_{i=j+1}^{L} \sigma_i^0 \right)  \ \ \ ; \ \ \ a \in \left\{ x, y  \right\} \ \ \ j \in \left\{ 1,2,\cdots, L  \right\}.  
\end{equation}
The $n=2$ subspace now consists of all products of the above operators.  After simplifying the products of single-site Pauli matrices we find that the $n=2$ subspace contains
\begin{eqnarray}
\left(\prod_{i=1}^{j-1} \sigma_i^0 \right) \sigma^a_j \left(\prod_{i=j+1}^{k-1} \sigma^z_i\right) \sigma^b_k \left(\prod_{i=k+1}^{L} \sigma_i^0 \right)    \ \ & ; & \ \ \  a,b \in \left\{ x, y  \right\}  \ \ \ j,k \in \left\{ 1,2,\cdots, L  \right\}.    \nonumber \\
\sigma^z_j\ \ \  & ; & \ \ \ \ j \in \left\{ 1,2,\cdots, L  \right\}
\end{eqnarray}
Note that, unlike the $n=1$ subspace, it is now identity matrices populating the chain to the left and right.  For $n=3$ we have that the subspace contains the operators 
\begin{eqnarray}
\left(\prod_{i=1}^{j-1} \sigma_i^z \right) \sigma^a_j \left(\prod_{i=j+1}^{k-1} \sigma^0_i\right) \sigma^b_k \left(\prod_{i=k+1}^{l-1} \sigma_i^z \right) \sigma^c_l \left(\prod_{i=l+1}^{L} \sigma_i^0 \right) \ \ & ; & \ \ \  a,b,c \in \left\{ x, y  \right\}  \ \ \ j,k,l \in \left\{ 1,2,\cdots, L  \right\}.    \nonumber \\
\left(\prod_{i=1}^{j-1} \sigma_i^z \right) \sigma^0_j \left(\prod_{i=j+1}^{k-1} \sigma^z_i\right) \sigma^a_k \left(\prod_{i=k+1}^{L} \sigma_i^0 \right)    \ \ & ; & \ \ \  a \in \left\{ x, y  \right\}  \ \ \ j,k \in \left\{ 1,2,\cdots, L  \right\}.     \nonumber \\
\left(\prod_{i=1}^{j-1} \sigma_i^z \right) \sigma^a_j \left(\prod_{i=j+1}^{k-1} \sigma^0_i\right) \sigma^z_k \left(\prod_{i=k+1}^{L} \sigma_i^0 \right)    \ \ & ; & \ \ \  a \in \left\{ x, y  \right\}  \ \ \ j,k \in \left\{ 1,2,\cdots, L  \right\}.   \nonumber
\end{eqnarray}
The first operator above has three operators either $\sigma^x$ or $\sigma^y$ with alternating insertions of strings of $\sigma^z$ and $\sigma^0$.  The second and third are an operator of the $n=1$ subspace with a $\sigma^z$ places somewhere on the right or a $\sigma^0$ somewhere on the left.  \\ \\
We can write the operators for the general subspace by introducing the shorthand
\begin{equation}
S_{kl}^{p,I} = \left(\prod_{i=k}^l \sigma^p_k\right) \left(\prod_i \sigma^z_{I(i)}\right) \ \ \ ; \ \ \ p \in \left\{  z, 0 \right\}
\end{equation}
The index $I$ contains all of the Pauli matrices between $k$ and $l$ that are not of type $p$.  In terms of these the subspaces are now
\begin{eqnarray}
& & S^{z,I_1}_{1,i_1} \sigma^{a_1}_{i_1+1}S^{0,I_2}_{i_1+2, i_2} \sigma^{a_2}_{i_2+1} S^{z,I_3}_{i_2+2,i_3} \cdots  \sigma^{a_k}_{i_k+1} S^{0,I_{k+1}}_{i_{k+2},L} \ \ \ ; \ \ \  n = k + 2\sum_{j=1}^{k+1} |I_j| \ \ \ ; \ k \ \textnormal{odd}   \nonumber \\
& & S^{0,I_1}_{1,i_1} \sigma^{a_1}_{i_1+1}S^{z,I_2}_{i_1+2, i_2} \sigma^{a_2}_{i_2+1} S^{0,I_3}_{i_2+2,i_3} \cdots  \sigma^{a_k}_{i_k+1} S^{0,I_{k+1}}_{i_{k+2},L} \ \ \ ; \ \ \  n = k + 2\sum_{j=1}^{k+1} |I_j| \ \ \ ; \ k \ \textnormal{even}    \nonumber
\end{eqnarray}
Note that, for even $n$ the operators on the left and right (with $|I_1| = |I_{k+1}| = 0$) is of the same type.  This provides another way in which the dramatic difference between the behavior of operators in periodic and open boundary conditions may be analysed and understood.  

\end{document}